\documentclass[a4paper]{jpconf}

\usepackage[utf8]{inputenc}
\usepackage[T1]{fontenc}
\usepackage[dvips]{graphicx}

\begin{document}

\title{Quasinormal modes for the charged Vaidya metric}
\author{Cecilia Chirenti $^1$ and Alberto Saa $^2$}

\address{$^1$ Centro de Matem\'atica, Computa\c c\~ao e Cogni\c c\~ao, UFABC, 09210-170 Santo
Andr\'e, SP, Brazil}

\address{$^2$ Departamento de Matem\'atica Aplicada, UNICAMP, 13083-859 Campinas, SP, Brazil}

\begin{abstract}
The scalar wave equation is considered in the background of
a charged Vaidya
metric in double null coordinates $(u,v)$ describing
a non-stationary charged black hole
with varying mass $m(v)$ and charge $q(v)$.
The resulting time-dependent quasinormal modes
are presented and analyzed. We show, in particular, that it is possible
to identify some signatures  in the quasinormal frequencies from
 the creation of  a naked singularity.
\end{abstract}

\section{Introduction}
Even though macroscopically charged black holes are not expected to be produced in realistic  astrophysical scenarios, we can obtain fascinating information on the structure and behavior of Einstein equations and general relativity with the study of these objects. Here, we used the charged Vaidya metric to model time-dependent black holes with an electric charge, and study the behavior of the quasinormal modes in a non-stationary background.

The charged Vaidya metric\cite{charge1,charge2,charge3} is an asymptotic flat and spherically symmetric solution of Einstein–Maxwell equations in the presence of a radial flow of a null-type fluid. For our purposes, it is more convenient to consider the charged Vaidya metric in double-null coordinates, as proposed by Waugh and Lake in \cite{uv}. In double-null coordinates
$(u,v,\theta,\phi)$, the charged Vaidya metric reads
\begin{equation}
ds^2 = -2f(u,v)dudv + r^2(u,v)d\Omega^2\,,
\label{metric}
\end{equation}
where $d\Omega^2$ stands for the metric on the unity sphere $(\theta,\phi)$.
The Einstein-Maxwell equations imply that the functions
  $f(u,v)$ and
$r(u,v)$ and $h(u,v)$ obey
\begin{equation}
f(u,v) = 2B(v)r_{,u}(u,v)
\label{f}
\end{equation}
and
\begin{equation}
r_{,v}(u,v) = -B(v)\left( 1 - \frac{2m(v)}{r}   +\frac{q^2(v)}{r^{2}} \right),
\label{r}
\end{equation}
where $B(v)$, $m(v)$ and $q(v)$ are arbitrary integration functions. With the choice of $B(v) = $ constant, one can interpret
  $m(v)$ and  $q(v)$, respectively, as the  total mass and the total charge of
  the solution. Both
functions are only required to be monotonic, but a further condition
must be imposed due to the weak energy condition in the case where
both $\dot{m}$ and $\dot{q}$ are non null \cite{Ori91}.

\begin{figure}[!htb]
\begin{minipage}[b]{3in}
\begin{center}
  \includegraphics[angle=270,width=0.8\linewidth]{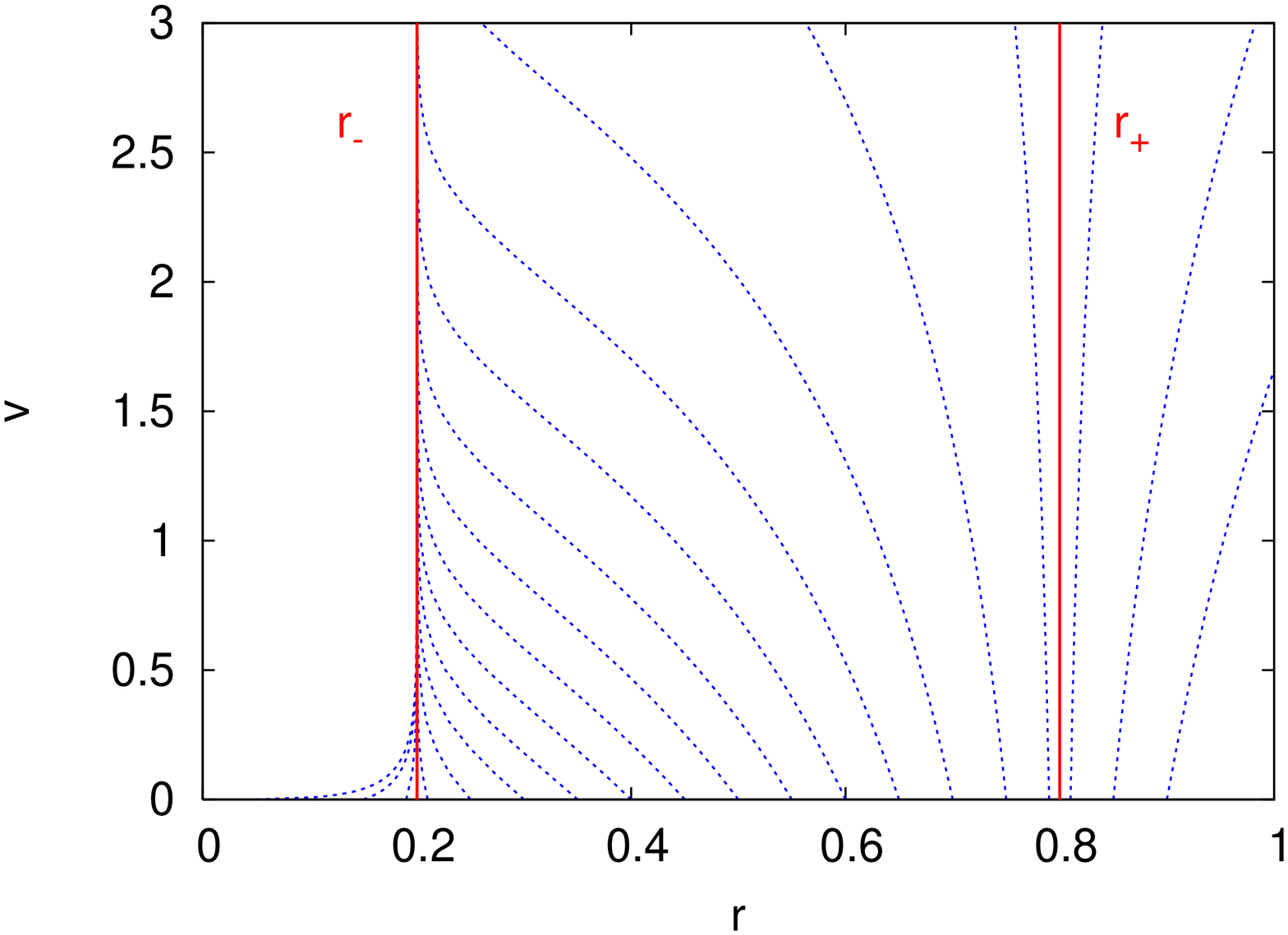}
\end{center}
\caption{Example of null-geodesics obtained from eq.~(\ref{r}) for
a RN black hole, with $B = -1/2$, $m = 0.5$, $q = 0.4$,
$n = 4$ and $\Lambda = 0$ and taking $r(u,v=0) = -u/2$. Here $r_+$ and
$r_-$ are defined as usual as $r_{\pm} = m \pm \sqrt{m^2 - q^2}$.}
\label{fig1}
\end{minipage}
\hspace{.2in}
\begin{minipage}[b]{3in}
\begin{center}
  \includegraphics[angle=270,width=0.8\linewidth]{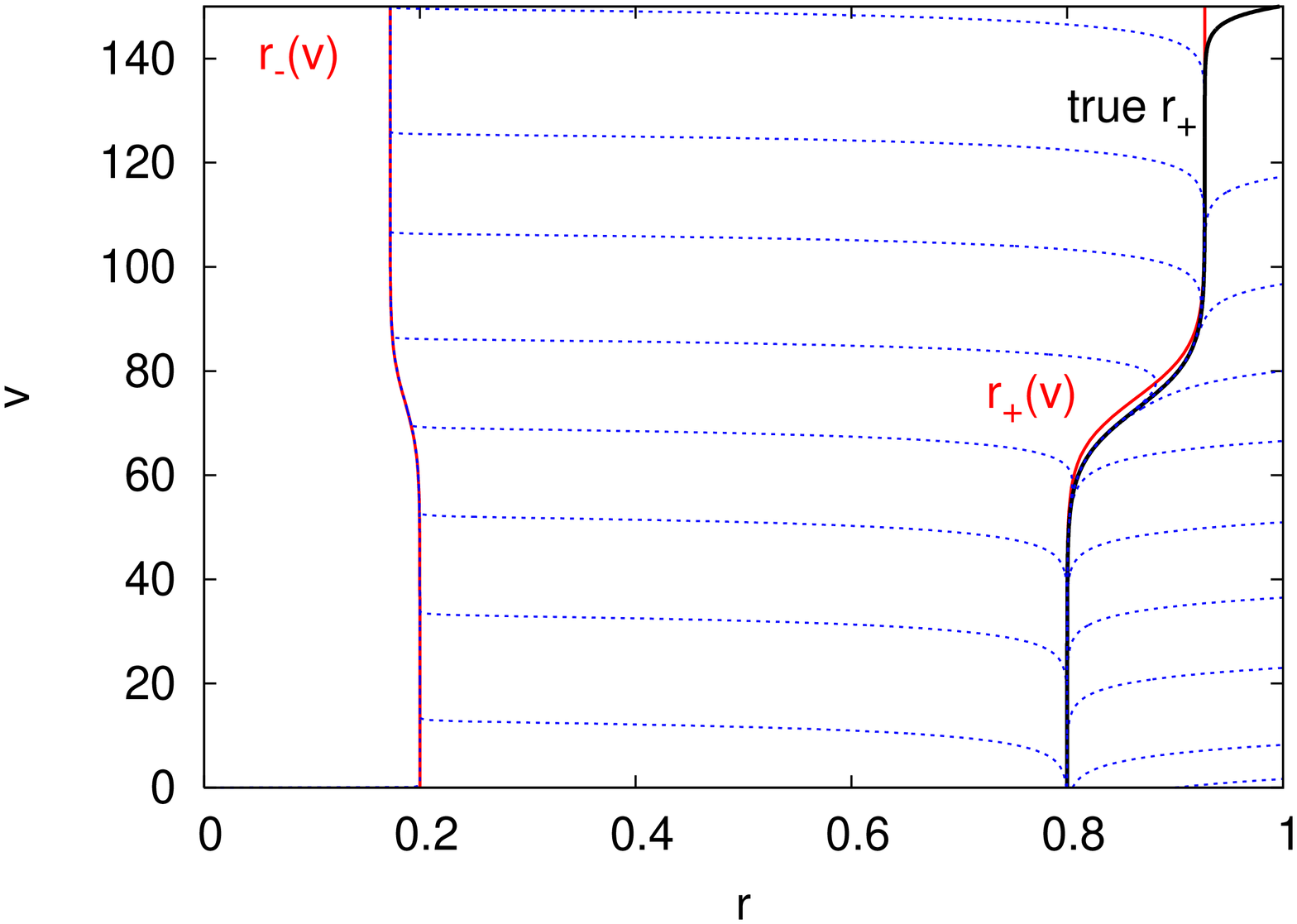}
\end{center}
\caption{Same as in Figure~\ref{fig1}, but this time for a time dependent charged black hole, with the choices $B = -1/2$, $m(v) = 1.0$ and $q(v) = \frac{1}{2}(q_2-q_1)[1 + \tanh
    \rho(v-v_m)] + q_1$,
with $q_1 = 0.5$, $q_2 = 0.7$, $\rho = 0.1$ and $v_m = 90$.}
\label{fig3}
\end{minipage}
\end{figure}

The solutions of Eq. (\ref{r}) corresponds to the congruence of ingoing and outgoing null geodesics of the metric (\ref{metric}). We adopt here the semi-analytical method proposed in \cite{SaaGirotto,Saa} to construct such geodesics.
In Fig.~ \ref{fig1} we present as an example the null geodesics obtained for
a case with $\dot{m} = \dot{q} = 0$, that is, a Reissner-Nordstr\"om (RN) black hole,
in this parametrization. Fig.~\ref{fig3} presents the same solutions for a time dependent background. For an interesting analysis of the time dependent horizons in the Vaidya metric, see \cite{Booth10}.

\section{Scalar perturbations}

The evolution of a massless scalar field in this background is given
by the Klein-Gordon equation:
\begin{equation}
\frac{1}{\sqrt{-g}}\left(\sqrt{-g}g^{\mu \nu}\Psi_{,\nu}\right)_{,\mu}
= 0\,.
\label{wave1}
\end{equation}
Decomposing the field $\Psi(u,v,\theta,\phi)$ in spherical harmonics
\begin{equation}
\Psi(u,v,\theta,\phi) =
\sum_{\ell,m}r^{-1}\varphi(u,v)Y_{\ell m}(\theta,\phi)\,,
\label{psi}
\end{equation}
and using eqs.~(\ref{f}) and (\ref{r}), we find
\begin{equation}
\varphi_{,uv} + V(u,v)\varphi = 0\,,
\label{onda}
\end{equation}
with the potential $V(u,v)$ given by
\begin{equation}
V(u,v) = \frac{1}{2}f(u,v)\bigg(\frac{\ell(\ell+1)}{r^2}
+ \frac{2m(v)}{r^{3}}
 - \frac{2q^2(v)}{ r^{4}}\bigg)\,.
\label{potential}
\end{equation}
For more details on the numerical algorithm, initial conditions and wave extraction process for Eq. (\ref{onda}) we refer the reader to refs.~\cite{Abdalla06,Abdalla07}.

\section{Numerical results}

We first tested our code for static Schwarzschild and Reissner-Nordström configurations, with very good agreement with values found in the literature \cite{Konoplya03,Konoplya02}.

\begin{figure}[!htb]
\begin{center}
  \includegraphics[angle=270,width=0.45\linewidth]{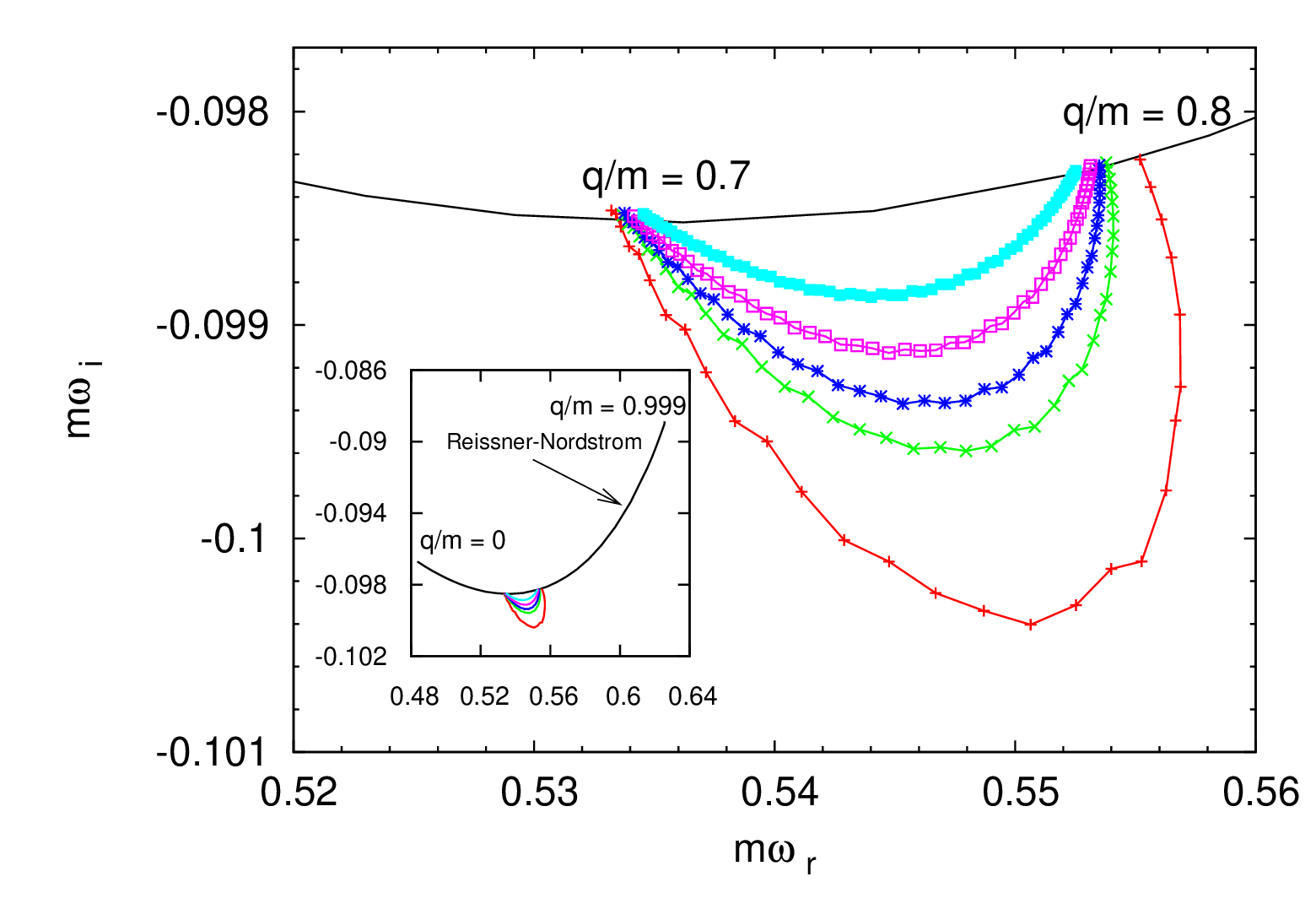}
  \includegraphics[angle=270,width=0.45\linewidth]{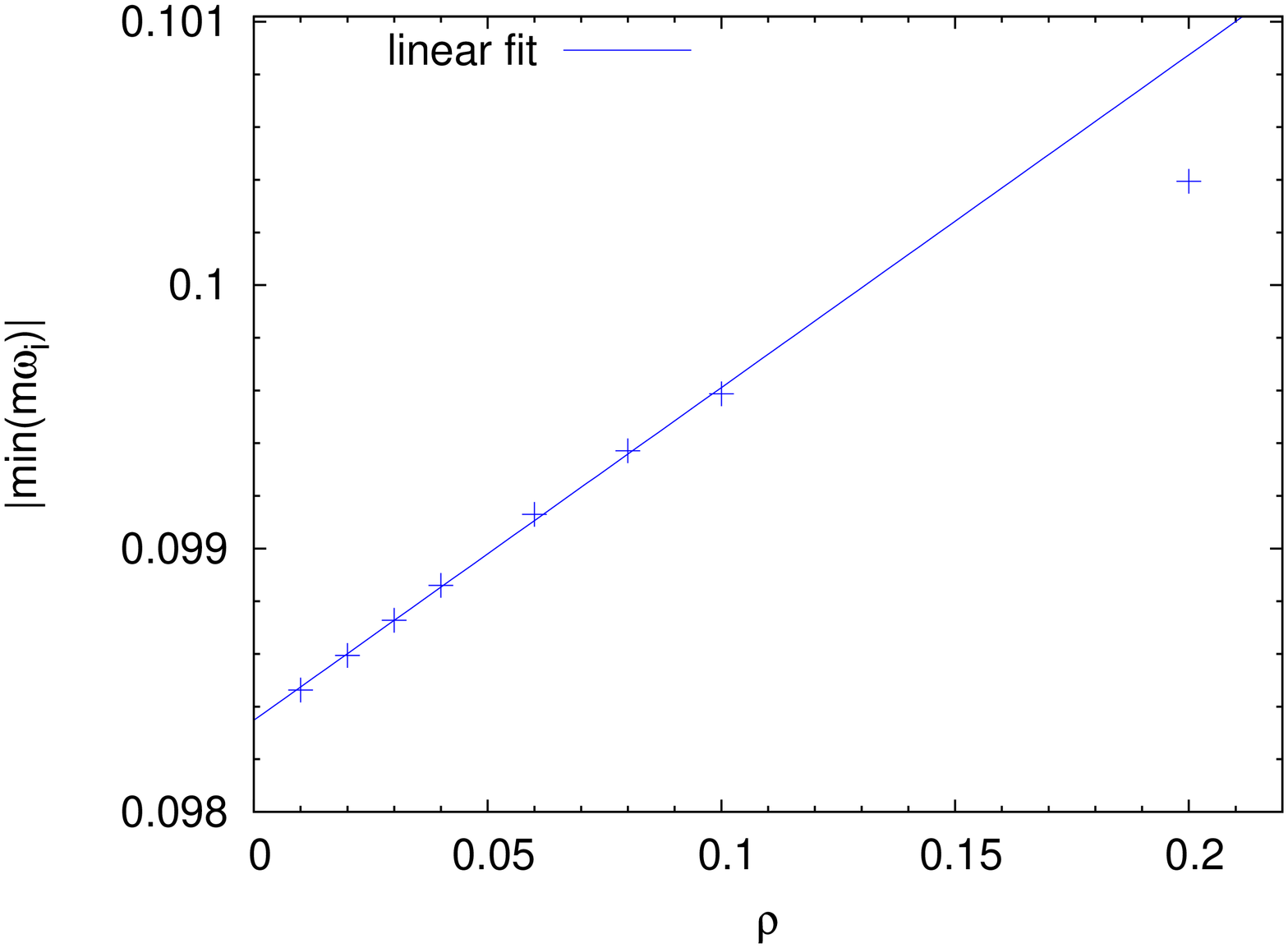}
\end{center}
\caption{Left plot:  quasinormal modes in the $\omega_r \times
  \omega_i$ plane, showing the transition between stationary
  states for the $\ell = 2$ scalar
  perturbation of a black hole with constant mass $m =
  0.5$ and $q(v) = \frac{1}{2}(q_2-q_1)[1 + \tanh
    \rho(v-v_m)] + q1$,
with $q_1 = 0.7$, $q_2 = 0.8$, $v_m = 75$ and different values of
$\rho$. Right plot: the minima of $\omega_i$ obtained in the transition shown in the left plot for different values of $\rho$, to quantify the non-adiabatic behavior.}
\label{wrxwi_01}
\end{figure}

In Fig.~(\ref{wrxwi_01}) we present the time variation in the $\omega_r \times
\omega_i$ plane of the real and
imaginary parts of the QNM of the scalar perturbation with $\ell = 2$
for a RN black hole with constant mass and a time dependent charge
function $q(v)$. 
The frequencies change in a similar way with the time variation of the
mass or the charge, but are typically more sensitive to the charge. The specific case depicted in
Fig.~(\ref{wrxwi_01}) corresponds to a charge variation
where $q/m$ goes from 0.7 to 0.8, where we see how the variation goes from one
stationary state to another through a non-adiabatic trajectory.
The relevant timescale here for characterizing mass or charge changes
which are adiabatic or not is essentially the same one of the $q=0$ case studied in \cite{Abdalla07}: QNM modes will relax and follow an adiabatic
 trajectory in the $\omega_r \times
\omega_i$ plane whenever $r''_+(v) < \omega_i (v)$ .

\begin{figure}[!htb]
\begin{center}
  \includegraphics[angle=270,width=0.45\linewidth]{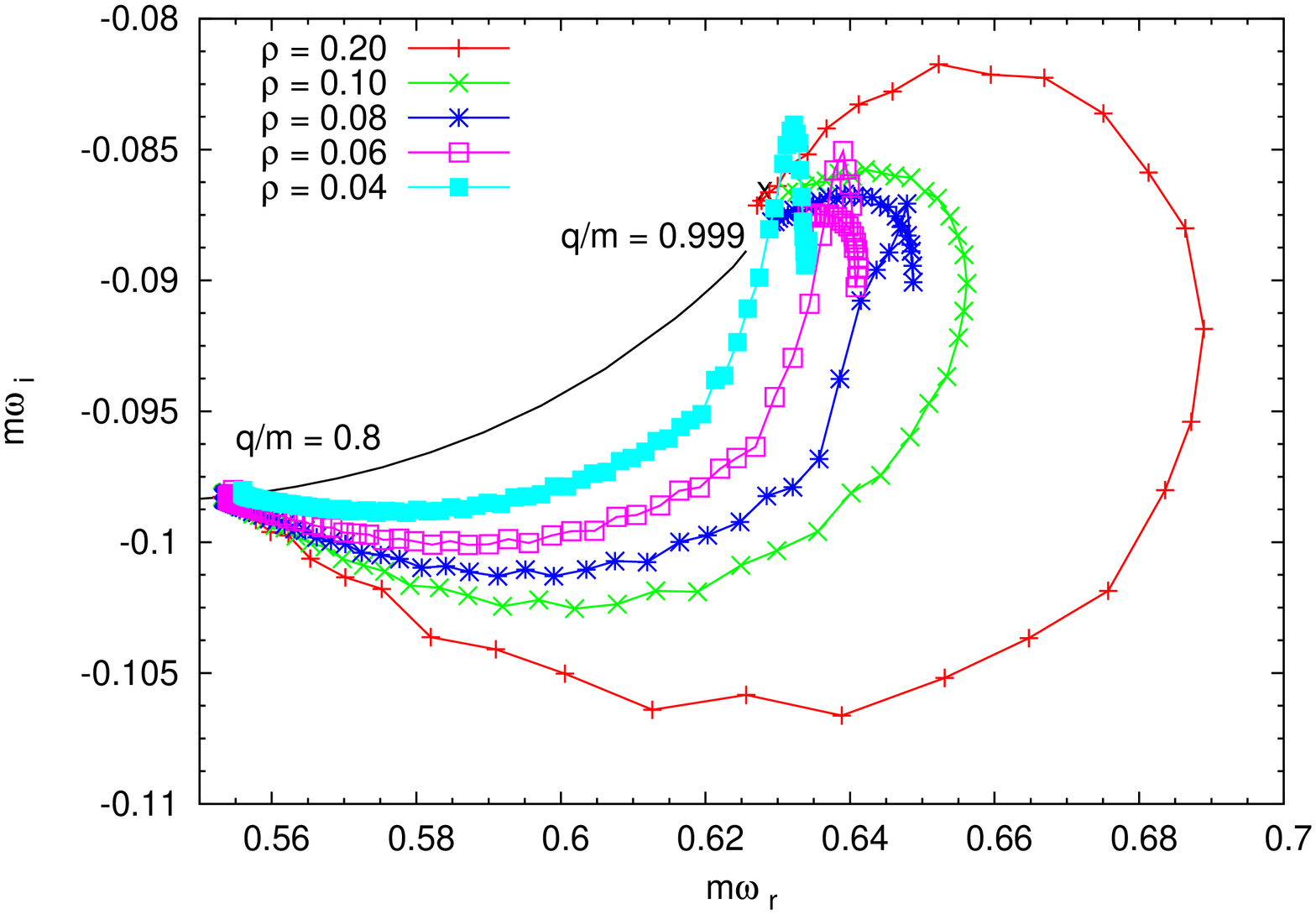}
  \includegraphics[angle=270,width=0.45\linewidth]{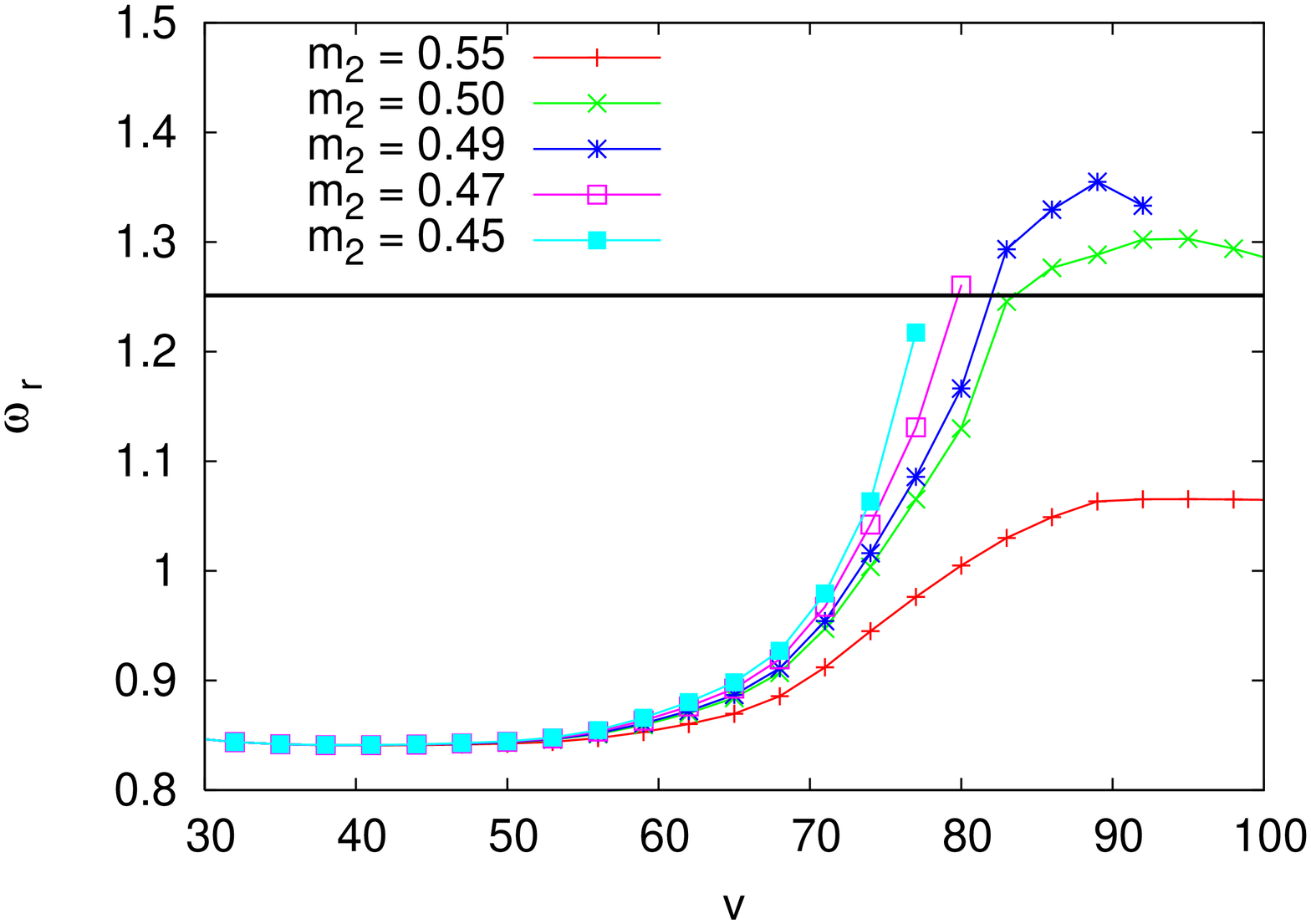}
\end{center}
\caption{Left plot: same as in the left plot shown in Fig.~(\ref{wrxwi_01}), but this time with $q_2 = 0.5$ so that the RN black hole becomes extreme as $v \to \infty$. Right plot: $\omega_r(v)$ of a scalar perturbation with $\ell = 2$ obtained for ``artificial'' background configurations with $m_2$ approaching and smaller than $q$; $\omega_i(v)$ presents a similar behavior.}
\label{wrxwi_04}
\end{figure}

We would like also to use our formalism and numerical setup to probe the
formation of a naked singularity in the spacetime. In order to do so
and avoid numerical problems as $q \to m$, we choose $m = m_0$ and
$q(v) = (q_2-q_1)[1 + \tanh \rho(v-v_m)] + q_1$, taking $q_2 =
m_0$. This implies that the RN black hole will be (rigorously) extremal
only for $v \to \infty$, but allows us to probe configurations very
near this end state at finite times. By varying the parameter $\rho$,
we can control the speed with which the black approaches the extremal
state.
The results are shown in Fig.~(\ref{wrxwi_04}). In the left plot we show how the results depend on the manner in
which $q(v) \to m$. We remind here that an extremal RN black hole is
not a naked singularity, and its QNMs are well defined as the appropriated limit of the non-extremal case \cite{Onozawa96,Berti04,Daghigh08}. The results in the right plot of Fig.~\ref{wrxwi_04} show what happens when we try to artificially create a naked singularity. We chose configurations with constant charge $q$ and time dependent mass functions  with final mass $m_2$ approaching and smaller than $q$. The results for $\omega_r$ are shown in the right plot of Fig.~\ref{wrxwi_04} and are truncated when $m(v) = q$ since the simulation crashes at this point, as expected.

\section{Final remarks}

Our results confirm, for the charged case, the same  non-stationary
behavior corresponding to some sort of inertia for QNM frequencies  identified
in \cite{Abdalla06}. In particular, for situations with
$r''_+(v) > \omega_i (v)$, the QNM frequencies will not follow in the $\omega_r \times
\omega_i$ plane a trajectory corresponding to the instantaneous frequency associated to a RN black hole of a given $q/m$ ratio, and a certain inertial behavior for the QNM can be identified. Moreover, we see that the faster the change in $m(v)$ or $q(v)$ is, the bigger the deviation from the stationary regime will be (see Fig. (3)). These points are now under investigation, and a more complete analysis will appear soon \cite{CeciliaSaa}.

As for the $q=0$ case considered in \cite{Abdalla06}, an interesting extension of this work would be analysis
of the highly damped QNM, the so-called overtones. Since
for such overtones the ratio $|\omega_i/\omega_r|$ is always larger than
for the fundamental ($n = 0$) QNM considered here, including, for sufficient
large $n$, cases for which $|\omega_i/\omega_r|>1$, it would be
interesting to check if the non-stationary inertial behavior
could somehow be attenuated for $n > 0$. We notice
that the numerical analysis presented here cannot be extended
directly to the $n > 0$ case since one cannot identify
the $n > 0$ frequencies with sufficient accuracy. We
believe this could be attained, in principle, by means of
the WKB approximation.

\ack
This work was supported by CNPq, FAPESP and the Max Planck Society. It is a pleasure to thank Rodrigo Panosso Macedo and Luciano Rezzolla for useful discussions.

\end{document}